# Negative refraction of a three-dimensional metallic photonic crystal

Ali Mahmoudi, Department of physics, Faculty of sciences, University of Qom, Iran & Institute for Advanced Studies in Basic Sciences (IASBS),Zanjan 45195,P. O. Box 45195-1159,Iran

Abbas Semnani, Department of Communication, Faculty of Electrical Engineering, K.N.Toosi University of Technology, Tehran, Iran

Ramazan Alizadeh, University of Isfahan, Isfahan, Iran

**Abstract** A metamaterial with a negative effective index of refraction is made from a three-dimensional hexagonal lattice photonic crystal with a metallic basis embedded in foam. It has been simulated with Ansoft HFSS$_{TM}$ in a frequency range from 7.0 to 12.0 GHz. Simulated results tested experimentally and negative refraction verified in some frequencies. Experimental results are in good agreement with simulations.

**Introduction** Photonic crystals (PCs) are structures with periodic arrangements of dielectrics or metals. In such medium, we can write this equation for magnetic field H:

$$\nabla \times \left( \frac{1}{\varepsilon(r)} \nabla \times \vec{H}(r) \right) = \left( \frac{\omega}{c} \right)^2 \vec{H}(r) \qquad (1)$$

This is an eigenvalue equation. It can be shown that $\nabla \times \left( \frac{1}{\varepsilon(r)} \nabla \times \right)$ is a Hermitian operator [1]. With periodic $\varepsilon(r)$ Eq.1 is similar to Schrödinger equation in solid state physics:

$$\varepsilon(\vec{r}+\vec{R}) = \varepsilon(\vec{r}) \qquad (2)$$

Where $\vec{R}$ is a vector for describing lattice symmetry, and is a linear combination of basic lattice vectors **a**, **b** and **c**:

$$R = l a + m b + n c \quad l,m,n = \pm 1, \pm 2, \pm 3 \qquad (3)$$

Such periodicity makes limitations on acceptable eigenvalue ($\frac{\omega}{c}$). These limitations are the origin of interesting properties of photonic crystals including photonic band gap, negative refraction and so on. Based on Bloch theorem, each component of H-filed must have this form:

$$H(r+R) = \exp(ik.R)H(r) \qquad (4)$$

H can be expanded in plane waves:

$$H(r) = \sum H_k e^{iK.r} \qquad (5)$$

There are several numerical methods for calculating H(r) and band structure - $\omega(k)$ - of a photonic crystal material. Finite difference time domain (FDTD) method, finite element method and plane wave expansion method are used commonly. FDTD and finite element methods are useful for calculating modes, plane wave expansion is a good method for calculating band structures. In HFSS, finite element method is used.

An effective index of refraction for the crystal is used to describe the overall reflectivity from the photonic structure:

$$n = c\frac{d\omega}{dk} \qquad (6)$$

Thus calculating band structure of a photonic crystal numerically led to calculation of n. In experimental point of view, n can be calculated by Snell law. Many studies, both numerical and experimental were made recently in this area. All-angle negative refraction from PCs was presented in [2] and negative refraction from metallic/dielectric PCs at frequency points was demonstrated numerically and experimentally in [3– 9 ]. Left-handed structures based on a dielectric photonic crystal (PC) with a negative refractive index [8] have been tested. Negative refraction demonstrated also in two dimensional metallic photonic crystals (PC) in microwave frequencies [9-10, 11].

Theoretical studies indicate that the underlying mechanism for the negative refraction in PCs is not unique. Notomi extensively studied light propagation and described different ways for achieving negative effective index of refraction $n_{eff}$ in strongly modulated two dimensional (2D) PCs [12].

Negative refraction may occur when the incident field couples to a band with convex equal frequency contours (EFCs) in **k**-space, where the conservation of the surface parallel component of the wavevector, **k**, combined with the "negative" curvature of the band causes the incident beam bend negatively.[12,13]. In this case, neither the group velocity nor the effective index is negative and the PC is essentially a positive index medium, exhibiting negative refraction. In another mechanism, the group velocity and the phase velocity derived from the band dispersion are antiparallel for all the values of **k**, leading to $n_{eff}$<0 for the PC.[14]. Both mechanisms are confirmed by recent experimental observations.[15,16,17] Negative refraction studies originate from left handed metamaterials[18,19] which are based on the proposal by Veselago,[20] composed from periodically arranged negative permeability $\varepsilon < 0$ and permittivity $\mu < 0$ metal components, providing an effective medium with $\sqrt{\varepsilon}\sqrt{\mu} < 0$.

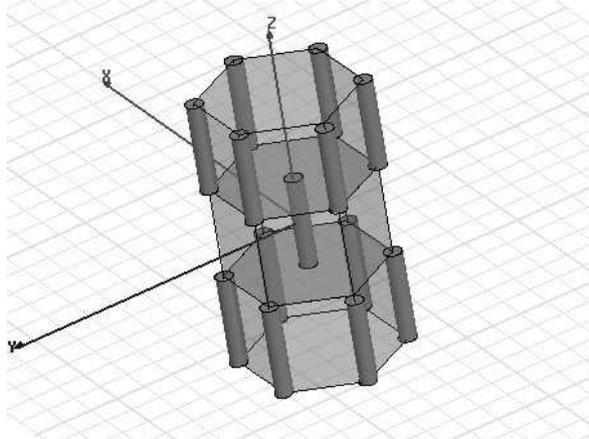

**Figure 1: Arrangement of copper bars in a hexagonal close-packed (hcp) lattice.**

## Design and Simulation

In the present work, we demonstrated negative refraction for a 3D structure made of metallic components. We have used Ansoft HFSS™ V.9 (High Frequency Structure Simulator) for simulating the refraction from the fabricated structure. HFSS is the industry-standard software for S-parameter and full-wave SPICE extraction and for the electromagnetic simulation of high-frequency and high-speed components. HFSS is widely used for the design of on-chip embedded passives, PCB interconnects, antennas, RF/microwave components, and high-frequency IC packages [23, 24]. Orientation of copper rods in lattice is shown in fig.1. HFSS model, its boundary conditions and excitations are shown in fig.2 and 3, respectively. Because the height of prism (40.5 cm) is larger than of 5 times of largest wavelength in simulation (4.3 cm), the system is as a semi-infinite structure. The refraction angles were predicted at 7.0– 12.0 GHz using Ansoft HFSS™. This frequency range is same as what tested for a similar 2D structure [8].

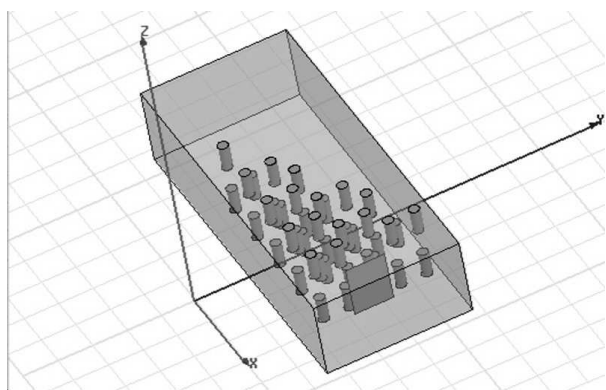

**Figure 2 : simulated structure in HFSS**

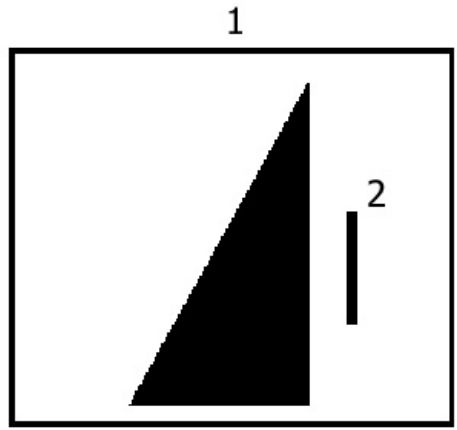

**Figure 3: Boundary conditions and excitations in HFSS model, on (1) radation boundary used, on (2) an incident wave excitation is applied.**

On the faces of box that surrounded the prism, we used radiation boundary condition; means that the problem solution is limited to this volume. The height of the cube in fig.2 is 4 times of the largest wavelength in simulations. An incident wave excitation with normal incidence defined on a rectangle (2) shown in figs.2, 3. This incident wave is a plane wave(TM or TE). For frequency steps 500 MHz ranging from 7 to 12 GHz we find radiation pattern in desired range of angles. The simulation results are shown in Figs.4, 5 and 6. Because of dimensions of cross section of the prism, the normal on the refraction surface is on $147°$ .Thus beams in lower angles are negative refracted beams. It is clear that negative refraction is occurred for 7.0, 9.5 and 10.5 GHz, as it can be seen from Figs.4 and 5. For these frequencies, negative refracted beam are strong and clear. For 11.5 GHz there are both positive and negative refractions with nearly equal intensities. In 12.0 GHz, the positive refracted beam is the dominant. The results are listed in Table.1.

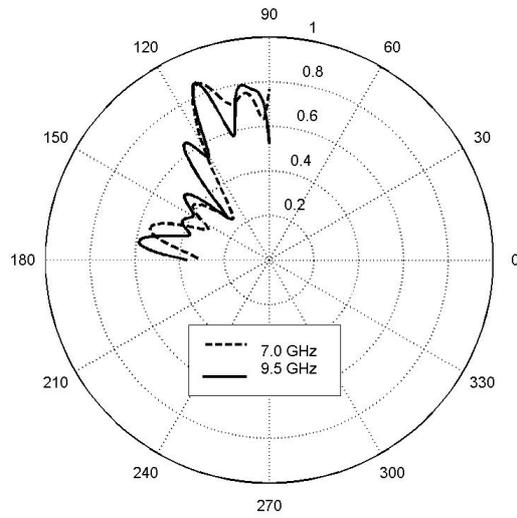

Figure 4 : calculated pattern of refracted E field for 7.0 and 9.5 GHz, negative refraction occurred for both frequencies.

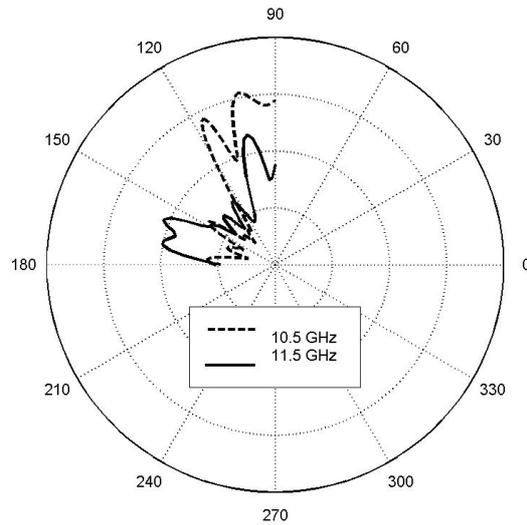

Figure 5: calculated pattern of refracted E field for 10.5 and 11.5 GHz, for 11.5 GHz both negative and positive refraction occurred, but in 10.5 GHz negative refraction occurred strongly.

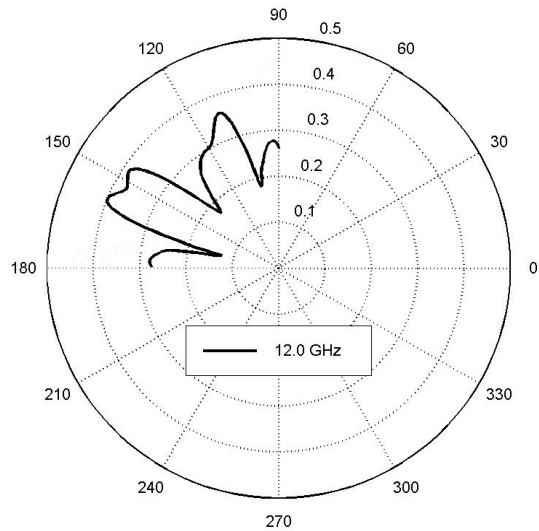

**Figure 6 : refracted pattern of E field for 12.0 GHz. Both negative and positive refraction are present.**

**Table 1: Negative and positive refractions.**

| Frequency (GHz) | Angle of negative refracted pick (deg) | Angle of positive refracted pick (deg) | Calculated n (refractive index) $n = \sin(\theta_r)/\sin(\theta_i)$ ($\theta_i = 32.4^0$) |
|---|---|---|---|
| 7.0 | 112 | No beam | -1.086 |
| 9.5 | 112.8 | No beam | -1.065 |
| 10.5 | 116 | Very weak | -0.977 |
| 11.5 | 102 | 158 | Two beams |
| 12.0 | 111 | 158 | Two beams |

## Experimental and Measurements

The 3D photonic crystal made of copper bars with diameter 0.63 cm and height of 1.5 cm arranged in a 3D array with a 3D hexagonal lattice. We draw the lattice on plates of foam (1.5 cm thick) and then insert copper bars into lattice points (Fig.7). The ratio of the radius *r* to the lattice constant *a(=b=c)* was *r /a*=0.2. We made this lattice layer by layer and the final structure was a prism with dimensions 12cm ×19cm × 40.5 cm.

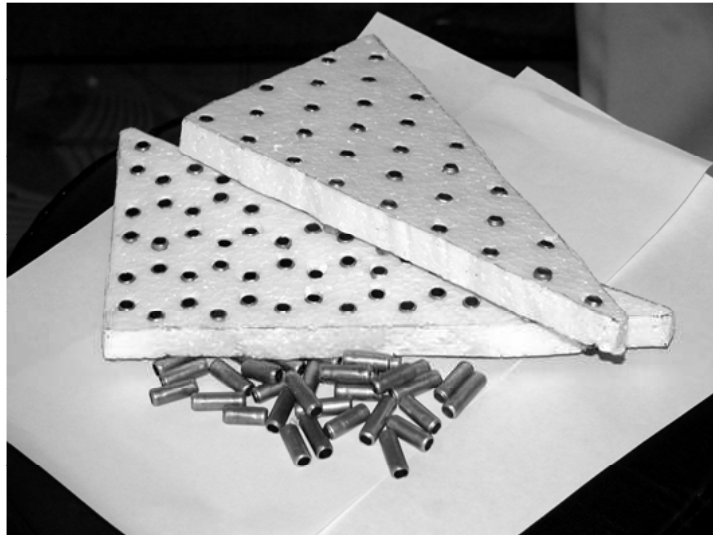

**Figure 7: Two principal layers that make 3D photonic crystal with a hexagonal closed packed lattice.**

Transmission measurements are performed to verify and test the negative refraction. The experimental setup consists of an HP 8510C network analyzer, a horn antenna as the transmitter, and a monopole antenna as the receiver. The experimental setup used in this study is similar to negative refraction studies in [4–6] and is illustrated in Figure 8. The radiating near field was measured at different frequencies for positive and negative refractions. Refraction experiments were performed in an anechoic chamber. A horn Antenna placed at 3.5 m from the PC acts as a plane-wave source (Fig. 8). Placing a piece of microwave absorber with a 12×12 cm² aperture in front of the PC narrows the incident beam. The angle of incidence $\theta = 0^0$ is chosen. On the other side of room another square horn attached to a goniometer swings around in two-degree steps to receive the emerging beam. Refraction is considered positive (negative) if the emerging signal is received to the left (right) of the normal to the surface of refraction of the PC.

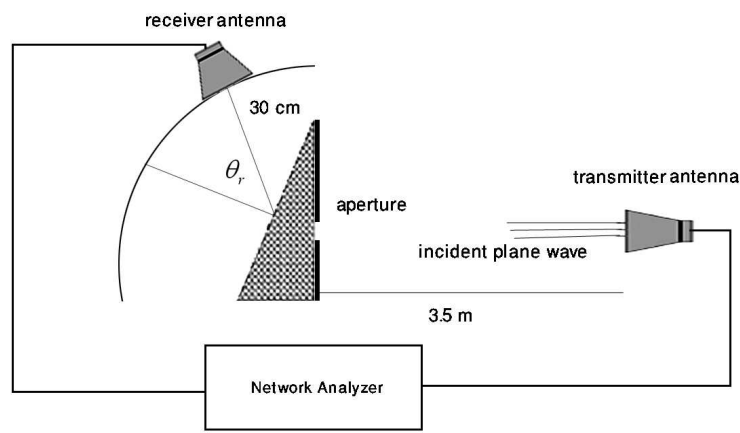

**Figure 8: Experimental setup for refraction test. Prism has dimensions 12cm ×19cm × 40.5 cm.**

In all the experimental and theoretical results, the electric field is kept parallel to the bars. Results (Figs.9, 10, 11, 12 and 13) show that a PC can exhibit negative refraction in a large frequency range. It is clear that in 7.0, 9.5 and 10.5 GHz we have only negative refraction, because the refracted wave centered at an angle lower than $147^0$. But in 11.5 and 12.0 GHz both positive and negative refraction occurred.

For avoiding diffraction effects, the structure must be fabricated precisely. The higher orders of diffraction eliminated by maintaining a low-frequency range where $\frac{a}{\lambda}$ is lower than 0.5 [21, 22]. Thus in our work the refraction in large wavelength limit (7 GHz) is not due to diffraction from fist row of copper bars. But in other end of the band, near 12 GHz it can be due to diffraction.

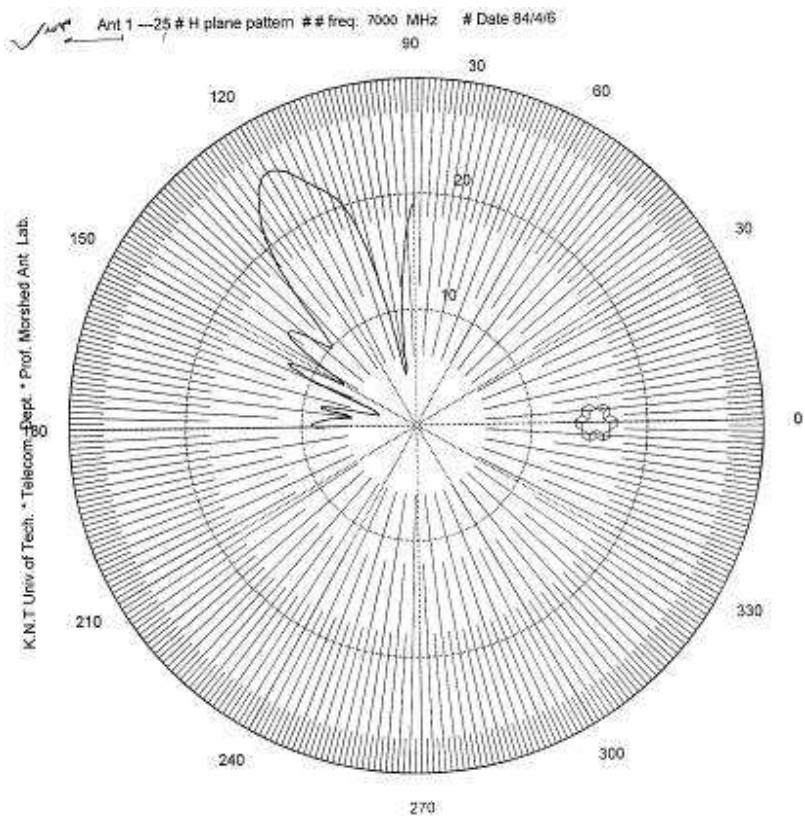

Figure 9 : measured pattern of refracted E filed for 7.0 GHz. Negative refraction occurred, pick of refracted wave is located on $120^0$ (normal direction is $147^0$).

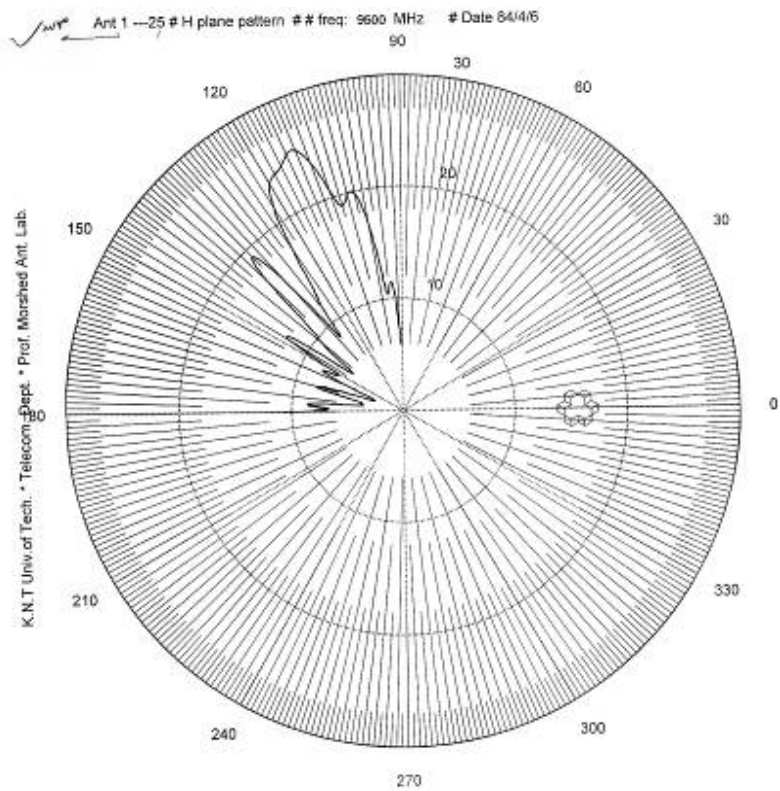

**Figure 10: measured refracted E field pattern at 9.5 GHz, negative refraction occurred.**

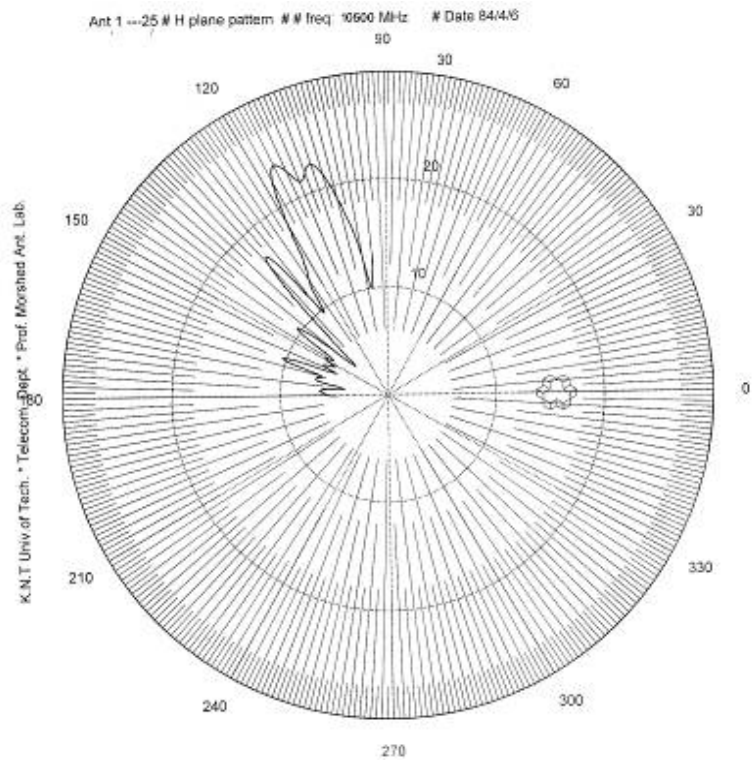

**Figure 11 : negative refraction for 10.5 GHz.**

It can be seen that with increasing of frequency, the level of the refracted waves decrease. In Figs.12 and 13, it is evident that negative and positive refractions have nearly equal level of intensity, but there are not same angles between them and the normal on the refraction surface. The results are listed in Table.2 for better comparison with simulation results gathered in Table.1.

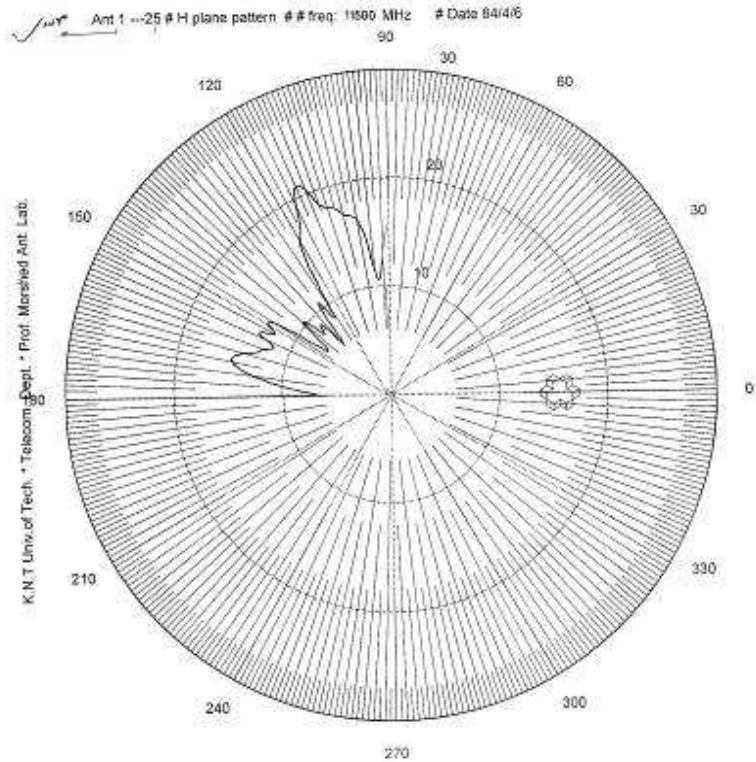

**Figure 12 : positive and negative refraction for 11.5 GHz. Positive pick located on angle 168° and negative pick located in angle 114$^0$.**

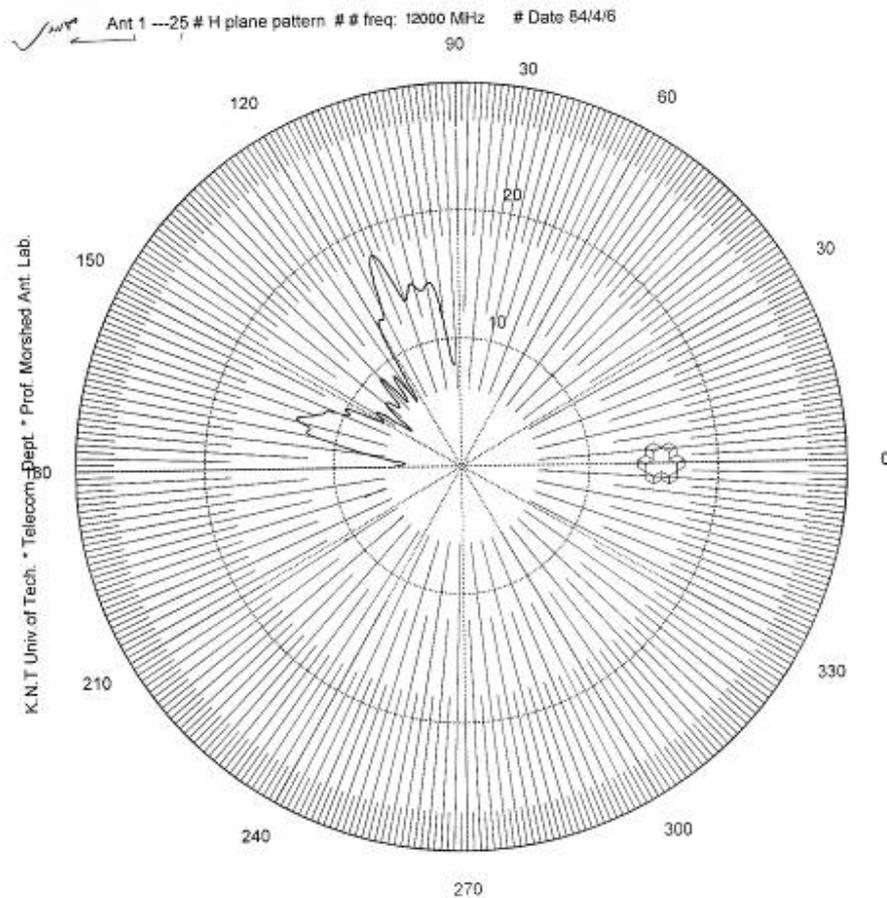

**Figure 11: : positive and negative refraction for 12 GHz. Positive pick located on angle $162^0$ and negative pick located in angle $112^0$**

Table 2: Measured results.

| Frequency (GHz) | Angular position of Negative refracted beam (deg) | Angular position of Positive refracted beam (deg) | Experimental n (refractive index) $n = \sin(\theta_r)/\sin(\theta_i)$ $(\theta_i = 32.4^0)$ |
|---|---|---|---|
| 7.0 | 120 | 155 | -0.87 |
| 9.5 | 113.5 | 149 | -1.046 |
| 10.5 | 116 | 160 | -0.977 |
| 11.5 | 113.5 | 167 | Two beams |
| 12.0 | 112 | 163 | Two beams |

In comparison to results reported in [9] for a 2D prism, in 7.0 GHz, they reported positive refraction, but here Negative refraction occurred. They observed no transmission between 7.1 and 8.3 GHz that is same as we observed. They observed that between 8.3 upto 11GHz, two signals emerged on the positive and negative sides of the normal. We observed similar results but it occurred from 11 GHz to higher frequencies. Between our results, the negatively refracted signal is strongest around 7 GHz, but they reported it 10.7 GHz. They also observed that with the increase in frequency, although both positively and negatively refracted signals are observed but positive signal gets weaker while negative signal gets stronger. Here we observed a different result, positive signal gets stronger and negative signal gets weaker with the increase in frequency.

The experimental refractive index *n* is obtained from applying Snell's law $n = \sin(\theta_r)/\sin(\theta_i)$ to each emerging beam. As it can be seen from Table.2, measured negative index of refraction at 10.5 GHz is exactly same as calculated result. The validity of Snell's law in metallic PCs. has been established earlier [21]. Comparison between Tables.1 and 2, shows that our calculations and measurements are in good agreement.

## Conclusions

We have demonstrated negative refraction in a simple 3D metallic photonic crystal. Simulation results and measurements are in good agreement. The results show that negative refraction occurred for different frequencies and in some frequency points there were both positive and negative refractions. The results of simulations and experimental measurements are in good agreement (Table.1 and Table.2). From scalability of Maxwell equation for H (Eq.1), we can expect same results for higher frequencies if we downscale dimensions of the structure appropriately.